\documentstyle[preprint,aps,prb]{revtex}
\def\be{\begin{equation}}
\def\ee{\end{equation}}
\begin{document}
\draft
\title{Boussinesq Solitary--Wave as a Multiple--Time Solution \\
of the Korteweg--de Vries Hierarchy}
\author{R. A. Kraenkel$^1$, M. A. Manna$^2$, J. C. Montero$^1$ and J. G.
Pereira$^2$ \cite{kra}}
\vskip 0.5cm
\address{$^1$Instituto de F\'{\i}sica Te\'orica\\
Universidade Estadual Paulista\\
Rua Pamplona 145\\
01405-900\, S\~ao Paulo -- Brazil \\
\vskip 1.0cm
$^2$Physique Math\'ematique et Th\'eorique, URA-CNRS 768\\
Universit\'e de Montpellier II\\
34095 Montpellier Cedex 05\\
France}
%\date{\today}
\maketitle
\begin{abstract}

We study the Boussinesq equation from the point of view of a multiple-time
reductive perturbation method. As a consequence of the elimination of the
secular producing terms through the use of the Korteweg--de Vries hierarchy, we
show that the solitary--wave of the Boussinesq equation is a solitary--wave
satisfying simultaneously all equations of the Korteweg--de Vries hierarchy,
each one in an appropriate slow time variable.

\end{abstract}


\vfill \eject
\section{Introduction}

As is well known, the Boussinesq model equation
\begin{equation}
u_{tt} - u_{xx} + u_{xxxx} - 3 (u^2)_{xx} = 0 \, ,
\label{1}
\ee
where $u(x,t)$ is a one--dimensional field and the subscripts denote partial
differentiation, is completely integrable. \cite{zakha} It is considered
as an intermediate long--wave equation since its long--wave limit with a
further restriction to waves moving in only one direction yields, at
the lowest order, the Korteweg--de Vries (KdV) equation
\be
u_t - 6 u u_x + u_{xxx} = 0 \, ,
\label{3}
\ee
which is an equation governing general weak nonlinear long--wave dynamics of
dispersive systems. \cite{calo} Equation (\ref{1}) has N--soliton solutions.
In particular, its solitary--wave solution is of the form~\cite{soli}
\be
u = -2 k^2 \, {\rm sech}^2 \left[k \left( x - \sqrt{1 - 4 k^2} \; t \right)
\right] \, ,
\label{2}
\ee
where $k$ is the wavenumber. Accordingly, its long--wave limit is related, also
at the lowest order,
to the solitary--wave solution of the KdV equation.

In a recent work, \cite{kmp} we have considered a perturbative scheme based on
the reductive perturbation method of Taniuti, \cite{taniu} modified by the
introduction of an infinite number of slow time-variables, which were given by
$\tau_3 = \epsilon^3 t$, $\tau_5 = \epsilon^5 t$, etc. Then, we have shown
that, as a consequence of a natural compatibility condition, a wave field
satisfying the KdV equation in the time $\tau_3$ must also satisfy all
equations of the KdV hierarchy, \cite{lax} each
one in a different slow time variable. The main reason for introducing these
time variables was that they allowed for the construction of a perturbative
series, valid for weak nonlinear dispersive systems, which was free of
solitary--wave related secularities.

As stated above, the solitary--wave  solution of the Boussinesq equation tends
to the KdV solitary--wave for small wave--numbers. What we will show here is
that, by making use of the perturbative scheme with multiple slow time--scales,
the solitary--wave of the Boussinesq equation may be written, in the slow
variables, as a solitary--wave solution to the whole set of equations of the
KdV hierarchy, each one in a different time--scale. This result follows both,
from the general long--wave perturbation theory, and from the observation that
the perturbative series truncates for a solitary--wave solution to the KdV
hierarchy equations, rendering thus an exact solution for the Boussinesq
equation.

The paper is organized as follows. In Section II the multiple--time formalism
is introduced for the Boussinesq equation, and the first few evolution
equations are obtained. In Section III we discuss how the
KdV hierarchy equations show up, and in Section IV we show how they
can be used to eliminate the solitary--wave related secularities of the
evolution
equations for the higher--order terms of the wave field.
In Section V, by returning from the slow to the laboratory coordinates, we
obtain the above mentioned relation between the solitary--waves of the
Boussinesq and the KdV hierarchy equations, namely that the solitary--wave of
the Boussinesq equation may be written, in the slow variables, as a
solitary--wave satisfying simultaneously all equations of the KdV hierarchy .
And finally, in Section VI, we summarize and discuss the results obtained.

\section{The Multiple Time Formalism}

In order to study the long-wave limit of eq.(\ref{1}), we will introduce slow
space and time variables based on the long--wave limit of the linear dispersion
relation
\be
\omega = k \left( 1 + k^2 \right)^{1/2} \, .
\label{6}
\ee
This limit corresponds to take
\be
k = \epsilon \kappa \, ,
\label{7}
\ee
with $\epsilon$ a small parameter. Expanding the dispersion relation (\ref{6}),
the solution of the corresponding linear Boussinesq equation is given simply by
\be
u = a \, {\rm exp} \, i \left[ \epsilon \kappa (x - t) - \frac{\epsilon^3
\kappa^3}{2}
 t + \frac{\epsilon^5 \kappa^5}{8} t - \frac{\epsilon^7 \kappa^7}{16} t +
\cdots \right] \, .
\label{9}
\ee
Based on this solution, we define now the slow space coordinate
\be
\xi = \epsilon (x - t) \, ,
\label{10}
\ee
and the infinite sequence of slow time coordinates
\be
\tau_3 = - \frac{\epsilon^3 t}{2} \quad ; \quad \tau_5 = \frac{\epsilon^5 t}
{8} \quad ; \quad \tau_7 = - \frac{\epsilon^7 t}{16} \quad ; \quad \cdots
\quad .
\label{11}
\ee
Accordingly, we have that
\be
\frac{\partial}{\partial x} = \epsilon \frac{\partial}{\partial \xi}
\label{12}
\ee
and
\be
\frac{\partial}{\partial t} = - \epsilon \frac{\partial}{\partial \xi} -
\frac{\epsilon^3}{2} \frac{\partial}{\partial \tau_3} + \frac{\epsilon^5}{8}
\frac{\partial}{\partial \tau_5} - \frac{\epsilon^7}{16}
\frac{\partial}{\partial \tau_7} + \cdots \quad .
\label{13}
\ee
Notice that in the definition of $\tau_{2n+1}$, we have already assumed
specific slow time normalizations, as inspired by the
long--wave expansion of the dispersion relation. As we are going to see, these
normalizations are exactly those necessary to cancel out the solitary--wave
related secularities appearing in the higher order evolution equations.

Returning to the nonlinear problem, we make now the expansion
\be
u = \epsilon^2 {\hat u} = \epsilon^2 \left(u_0 + \epsilon^2 u_2 + \epsilon^4
u_4 + \cdots \right) \, ,
\label{14}
\ee
and we suppose that $u_{2n} = u_{2n}(\xi , \tau_3 ,\tau_5 , ...),
n=0,1,2,\dots$, which corresponds to an extention in the sense of Sandri.
\cite{sandri} Substituting it, together with eqs.(\ref{12}) and (\ref{13}),
into the Boussinesq equation (\ref{1}), the resulting expression, up to terms
of order $\epsilon^4$, is:
\begin{eqnarray}
\left[ \frac{\partial^2}{\partial \xi \partial \tau_3} + \frac{\partial^4}
{\partial \xi^4} + \frac{\epsilon^2}{4} \left( \frac{\partial^2}{\partial
{\tau_3}^2} - \frac{\partial^2}{\partial \xi \partial \tau_5} \right) \right.
&+& \left.
\frac{\epsilon^4}{8} \left(\frac{\partial^2}{\partial \xi \partial \tau_7} -
\frac{\partial^2}{\partial \tau_3 \partial \tau_5} \right) + \cdots \right]
{\hat u}  \nonumber \\
- 3 \frac{\partial^2}{\partial {\xi}^2} \left[ (u_0)^2 \right. &+& \left. 2
\epsilon^2 u_0 u_2
+ \epsilon^4 (2 u_0 u_4 + (u_2)^2) + \cdots \right] = 0 \, .
\label{15}
\end{eqnarray}

We now proceed to an order--by--order analysis of the problem. At order
$\epsilon^0$ we get
\be
\frac{\partial}{\partial \xi} \left[ \frac{\partial u_0}{\partial \tau_3} - 3
\frac{\partial}{\partial \xi} (u_0)^2 + \frac{\partial^3 u_0}{\partial {\xi}^3}
\right] = 0 \, .
\label{16}
\ee
Integrating once and assuming a vanishing integration constant, we obtain
\be
\frac{\partial u_0}{\partial \tau_3} - 6 u_0 \, \frac{\partial u_0}{\partial
\xi} + \frac{\partial^3 u_0}{\partial {\xi}^3} = 0 \, ,
\label{17}
\ee
which is the KdV equation.

At order $\epsilon^2$, eq.(\ref{15}) yields
\be
\frac{\partial}{\partial \xi} \left[\frac{\partial u_2}{\partial \tau_3} - 6
\frac{\partial}{\partial \xi} (u_0 u_2) + \frac{\partial u_2}{\partial {\xi}^3}
\right] = \frac{1}{4} \frac{\partial^2 u_0}{\partial \xi \partial \tau_5} -
\frac{1}{4} \frac{\partial^2 u_0}{\partial {\tau_3}^2} \, .
\label{18}
\ee
Using eq.(\ref{17}), integrating once in $\xi$ and assuming a vanishing
integration constant, we obtain
\be
\frac{\partial u_2}{\partial \tau_3} - 6 \frac{\partial}{\partial \xi} (u_0
u_2) + \frac{\partial^3 u_2}{\partial {\xi}^3} = \frac{1}{4} \frac{\partial
u_0}{\partial \tau_5} - \frac{1}{4} \frac{\partial^5 u_0}{\partial {\xi}^5} + 3
u_0 \frac{\partial^3 u_
0}{\partial {\xi}^3} + \frac{9}{2} \frac{\partial u_0}{\partial \xi}
\frac{\partial^2 u_0}{\partial {\xi}^2} - 9 (u_0)^{2} \frac{\partial
u_0}{\partial \xi} \quad .
\label{19}
\ee
Equation (\ref{19}), as it stands, presents two problems. First, as
$u_{0\tau_5}$ is not known {\it a priori}, it cannot be solved for $u_2$. In
the next section we will show how to obtain the evolution of $u_0$ in the time
$\tau_5$ independently. The second
problem is that the term $(\partial^5 u_{0} / \partial {\xi}^5 )$, as a source
term for $u_2$, is a secular producing term when $u_0$ is chosen to be a
solitary--wave solution of the KdV equation. For instance, if we take the
solution of eq.(\ref{17}) proportional to $[{\rm sech^2} \, \theta]$, then
$(\partial^5 u_{0} / \partial {\xi}^5 )$ will contain a term proportional to
$[{\rm sech^2} \, \theta \; {\rm tanh} \, \theta]$. Being a solution of the
homogeneous part of eq.(\ref{19}), this term produces a resonance, giving rise
to non--uniformities in the perturbative series. It will turn out, however,
that $u_{0\tau_5}$ can be used to cancel out this secular term.

\section{The Rise of the Korteweg--de Vries Hierarchy}

As we have seen, the field $u_0$ satisfies the KdV equation in the time
$\tau_3$:
\be
u_{0\tau_3} = - u_{0\xi\xi\xi} + 6 u_0 \, u_{0\xi} \equiv F_3 \, .
\label{20}
\ee
The evolution of the same field $u_0$ in any of the higher--order times
$\tau_{2n+1}$ can then be obtained in the following way. \cite{kmp} First, to
have a well ordered perturbative scheme we impose that each one of the
equations
\be
u_{0\tau_{2n+1}} = F_{2n+1}(u_0, u_{0\xi}, \dots)
\label{20a}
\ee
be $\epsilon$--independent when passing from the slow $(u_0, \xi, \tau_{2n+1})$
to the laboratory coordinates $(u, x, t)$. This step selects all possible terms
to appear in $F_{2n+1}(u_0, u_{0\xi}, \dots)$. For instance, the evolution of
$u_0$ in time $\tau_5$ is restricted to be of the form
\be
u_{0\tau_5} = \alpha u_{0(5\xi )} + \beta u_0 u_{0\xi \xi \xi } + (\beta +
\gamma) u_{0\xi }u_{0\xi \xi } + \delta u_0^2 u_{0\xi } \, ,
\label{extra}
\ee
where $\alpha$, $\beta$, $\gamma$ and $\delta$ are constants. Then, by imposing
the natural (in the multiple time formalism) compatibility condition \cite{kmp}
\be
\Big( u_{0 \tau_3} \Big)_{\tau_{2n+1}} = \Big( u_{0\tau_{2n+1}} \Big)_{\tau_3}
\, ,
\label{extra1}
\ee
or equivalently,
\be
\left( {F_3} \right)_{\tau_{2n+1}} = \left( F_{2n+1} \right)_{\tau_{3}} \, ,
\label{21}
\ee
with $F_3$ given by eq.(\ref{20}), it is possible to determine any $F_{2n+1}$.
As it can be verified, \cite{kmp} the resulting equations are those given by
the KdV hierarchy. In particular, for $u_{0\tau_5}$ and $u_{0\tau_7}$ we
obtain respectively
\be
u_{0\tau_5} = u_{0(5\xi)} - 10 u_0 u_{0\xi\xi\xi} - 20 u_{0\xi} u_{0\xi\xi} +
30
(u_{0})^2 u_{0\xi} \, ,
\label{22}
\ee
and
\begin{eqnarray}
u_{0\tau_7} = &-& u_{0(7\xi)} + 14 u_0 u_{0(5\xi)} + 42 u_{0\xi} u_{0(4\xi)} +
\nonumber \\
&+& 70 u_{0\xi\xi} u_{0\xi\xi\xi} - 280 u_0 u_{0\xi} u_{0\xi\xi} -
70 (u_{0\xi})^3 - 70 (u_{0})^2 u_{0\xi\xi\xi} \, .
\label{23}
\end{eqnarray}
In principle, one could have an arbitrary constant multiplying the right-hand
side  of eqs.(\ref{22}) and (\ref{23}), which would correspond to an
arbitrarity in the slow time normalizations. However, as we will see in the
next section, the definition of the slow time variables we took implies that
these constants must be chosen to be one, since in this case the perturbation
theory is automatically rendered free of secularities. This choice is also the
one which makes the linear limit of the perturbation theory compatible with the
linear theory coming directly from eq.(\ref{1}).

\section{Higher Order Evolution Equations}

{}From this point on, we are going to consider some specific solutions to our
equations. First of all, we assume the solution of the KdV equation (\ref{20})
to be the solitary--wave solution
\be
u_0 = - 2 \kappa^2 {\rm sech^2} \left[\kappa (\xi - 4 \kappa^2 \tau_3 ) +
\theta \right] \, ,
\label{24}
\ee
where $\theta$ is a phase.
Moreover, since $u_0$ must satisfy also the equations of the KdV hierarchy, we
assume that $u_0$ given by (\ref{24}) be also a solitary--wave solution to all
equations of the KdV hierarchy, each one in a different slow--time variable.
This means that $u_0$ is actually
\be
u_0 = - 2 \kappa^2 {\rm sech^2} \left[\kappa \xi - 4 \kappa^3 \tau_3  + 16
\kappa^5 \tau_5 - 64 \kappa^7 \tau_7 + \cdots \right] \, .
\label{25}
\ee

We return now to eq.(\ref{19}) for $u_2$. Substituting $u_{0\tau_5}$ from
eq.(\ref{22}), we obtain
\be
u_{2\tau_3} - 6 (u_0 u_2)_{\xi} + u_{2\xi\xi\xi} = \frac{1}{2}\left[ - 3
(u_0)^2
u_{0\xi} + u_0 u_{0\xi\xi\xi} - u_{0\xi} u_{0\xi\xi} \right] \, .
\label{26}
\ee
In passing we notice that the use of the KdV hierarchy equation to express
$u_{0\tau_5}$ automatically canceled out the secular--producing term
$u_{0(5\xi)}$. Moreover,
using the solitary--wave solution (\ref{25}) for $u_0$, we see that the
right--hand side of eq.(\ref{26}) vanishes, leading to
\be
u_{2\tau_3} - 6 (u_0 u_2)_{\xi} + u_{2\xi\xi\xi} = 0 \, ,
\label{27}
\ee
which is the linearized KdV equation. We will assume for it the trivial
solution
\be
u_2 = 0 \, .
\label{28}
\ee
With this result, order $\epsilon^2$ is solved for the particular case we
chose.

At order $\epsilon^4$, and already assuming that $u_2 =0$, eq.(\ref{15}) gives
\be
{u_4}_{\tau_3 \xi} - 6 (u_0 u_4)_{\xi \xi} + {u_4}_{(4\xi)} = \frac{1}{8}
\left[-
{u_0}_{\tau_7 \xi} + {u_0}_{\tau_3 \tau_5} \right] \, .
\label{29}
\ee
Using equations (\ref{20}) and (\ref{22}) to express $u_{0\tau_3}$ and
$u_{0\tau_5}$ respectively, and integrating once in $\xi$, we obtain
\begin{eqnarray}
u_{4\tau_3} &-& 6 (u_0 u_4)_{\xi} + u_{4\xi\xi\xi} = \frac{1}{8}
\Big[ - u_{0\tau_7} - u_{0(7\xi)} + 16 u_0 u_{0(5\xi)} - 90 (u_0)^2
u_{0\xi\xi\xi} + \nonumber \\
&+& 70 u_{0\xi\xi} u_{0\xi\xi\xi} + 40 u_{0\xi} u_{0(4\xi)} - 300 u_0
u_{0\xi} u_{0\xi\xi} + 180 (u_0)^3 u_{0\xi} - 60 (u_{0\xi})^3 \Big] \, .
\label{30}
\end{eqnarray}
The term $u_{0(7\xi)}$ is the only resonant, that is, secular
producing term to the solution $u_4$.
Then, in the very same way we did before, we use the KdV hierarchy equation
(\ref{23}) to express $u_{0\tau_7}$. After we do that, the secular producing
term is automatically canceled out, and eq.(\ref{30}) becomes
\begin{eqnarray}
u_{4\tau_3} - 6 (u_0 u_4)_{\xi} + u_{4\xi\xi\xi} =
&-& 2 \big[ u_{0\xi} u_{0(4\xi)} - u_0 u_{0(5\xi)} + \nonumber \\
&+& 10  u_0 u_{0\xi}
u_{0\xi\xi} - 5 (u_{0\xi})^3 +
10 (u_0)^2 u_{0\xi\xi\xi} - 20 (u_0)^3 u_{0\xi} \big] \, .
\label{31}
\end{eqnarray}
Substituting again the solitary--wave solution (\ref{25}) for $u_0$, we can
easily
see that the nonhomogeneous term of eq.(\ref{31}) vanishes, leading to
\be
u_{4\tau_3} - 6 (u_0 u_4)_{\xi} + u_{4\xi\xi\xi} = 0 \, .
\label{32}
\ee
And again, we take the trivial solution
\be
u_4 = 0 \, .
\label{33}
\ee
It is easy to see that this is a general result that will repeat at any higher
order: for $n \geq 1$, the evolution of $u_{2n}$ in the time $\tau_3$,
after using the KdV hierarchy equation to express $u_{0\tau_{2n+1}}$ and then
substituting the solitary--wave solution (\ref{25}) for $u_0$, is given by a
homogeneous linearized KdV equation. Consequently, the solution
\be
u_{2n} = 0 \quad ; \quad n \geq 1
\label{34}
\ee
can be assumed for any higher order.

\section{Back to the Laboratory Coordinates}

Let us now take the solitary--wave solution to all equations of the KdV
hierarchy,
\be
u_0 = - 2 \kappa^2 {\rm sech^2} \left[\kappa \xi - 4 \kappa^3 \tau_3  + 16
\kappa^5 \tau_5 - 64 \kappa^7 \tau_7 + \cdots \right] \, ,
\label{35}
\ee
and rewrite it in the laboratory coordinates. First, recall that we have made
the expansion
\be
u = \epsilon^2 {\hat u} = \epsilon^2 \left(u_0 + \epsilon^2 u_2 + \epsilon^4
u_4 + \cdots \right) \, .
\label{36}
\ee
Thereafter, we have found a particular solution in which
\be
u_{2n} = 0 \quad ; \quad n \geq 1  \, .
\label{37}
\ee
Consequently, expansion (\ref{36}) truncates leading to an exact solution of
the form
\be
u = \epsilon^{2} u_0 \, ,
\label{38}
\ee
with $u_0$ given by eq.(\ref{35}).
Moreover, from eq.(\ref{7}) we see that the wave--number $\kappa$ is written in
terms of the corresponding laboratory one by
\be
\kappa = \epsilon^{-1} k \, .
\label{39}
\ee
Finally, the slow coordinates $\xi$ and $\tau_{2n+1}$ are related to the
laboratory ones, $x$ and $t$, according to eqs.(\ref{10}) and (\ref{11}). Then,
in the laboratory coordinates, the exact solution (\ref{38}) is written as
\be
u = - 2 k^2 {\rm sech^2} \, k \left[x - \left( 1 - 2 k^2 - 2 k^4 - 4 k^6 +
\cdots
\right) t \right] \, .
\label{40}
\ee
Now, the series appearing inside the parenthesis can be summed, with the result
\be
 1 - 2 k^2 - 2 k^4 - 4 k^6 + \cdots = \left(1 - 4 k^2 \right)^{1/2} \, \, .
\label{41}
\ee
Therefore, we get
\be
u = - 2 k^2 {\rm sech^2} \left[ k \left( x - \sqrt{1 - 4 k^2} \; t \right)
\right] \, ,
\label{42}
\ee
which is the well known solitary--wave solution of the Boussinesq
equation~(\ref{1}).

\section{Final Comments}

By applying a multiple time version of the reductive perturbation method of
Taniuti~\cite{taniu} to the Boussinesq model equation, and by eliminating
the solitary--wave related secular producing terms through the use of the KdV
hierarchy equations, \cite{kmp} we have suceeded in establishing a relation
between the solitary--wave satisfying all the equations of the KdV hierarchy
and that of the Boussinesq equation. In other words, we have shown that the
solitary--wave of the Boussinesq equation is given, in slow variables, by the
solitary--wave satisfying simultaneously all the equations of the KdV
hierarchy. Accordingly, while the KdV solitary--wave depends only on one slow
time variable, namely $\tau_3$, the solitary--wave of the Boussinesq equation
can be thought as depending on the infinite sequence of slow time variables.

The above considerations put in evidence the universal character played not
only by KdV, but by all equations of the KdV
hierarchy in relation to general weak nonlinear dispersive systems.
For such systems, as we have already said, it is always possible to define slow
variables in which the KdV equation emerges at the lowest relevant order of the
reductive perturbation method, and consequently, compatibility and
secularity--free requirements imply that all equations of the
KdV hierarchy emerge as well. In the case where the perturbative series
truncates, we may then obtain an exact solution of the original equation, which
is, in this sense, reconstructed from the perturbative expansion. Of course,
the Boussinesq solitary--wave is a well-known solution, but not the least, it
remains that a possible method of construction of solutions for more involved
system of  equations  can be envisaged. The return to the laboratory
coordinates, then, make the connection between a solution of the KdV hierarchy
and that of the equations governing the original system.

To conclude, we may conjecture that, whenever the original model equation has
an exact solitary--wave solution, the series may somehow be truncated
(or eventually summed) and a
relation will be established between the solitary--wave of the KdV hierarchy
and that of the original equation. On the other hand, when the original
nonlinear dispersive
system does not present an exact solitary--wave solution, the series will not
truncate. In this case, a secular--free expansion can still be obtained and
the process of returning to the laboratory coordinates can be made
order--by--order at any higher order, implying in a sucessive solitary--wave
velocity renormalization. \cite{kmp,kota}

\vspace{1 cm}
\section*{Acknowledgements}

The authors would like to thank CNPq-Brazil, CAPES-Brazil and FAPESP-Brazil for
financial support.

\end{document}